\documentclass[aps,prl,reprint,showpacs,showkeys,superscriptaddress,preprintnumbers,nofootinbib]{revtex4-1}
\usepackage[parfill]{parskip}    
\usepackage{graphicx}
\usepackage{amssymb}
\usepackage{epstopdf}
\usepackage{verbatim}
\usepackage{url}
\usepackage{hyperref}
\usepackage{subfigure}

\newcommand{\be}{\begin{equation}}
\newcommand{\bea}{\begin{eqnarray}}
\newcommand{\beq}[1]{\begin{equation}\label{#1}}
\newcommand{\eq}[1]{Eq.~(\ref{#1})}
\newcommand{\eqs}[2]{Eqs.\ (\ref{#1}) and (\ref{#2})}
\newcommand{\ee}{\end{equation}}
\newcommand{\eea}{\end{eqnarray}}
\newcommand{\eeq}{\end{equation}}
\newcommand{\lsim}{\!\mathrel{\hbox{\rlap{\lower.55ex \hbox{$\sim$}} \kern-.34em \raise.4ex \hbox{$<$}}}}
\newcommand{\gsim}{\!\mathrel{\hbox{\rlap{\lower.55ex \hbox{$\sim$}} \kern-.34em \raise.4ex \hbox{$>$}}}}

\newcommand{\vev}[1]{\langle #1 \rangle}
\newcommand{\abs}[1]{\left| #1 \right|}

\usepackage{color}

\begin{document}

\setlength{\baselineskip}{0.22in}

\preprint{MCTP-13-26}

\title{Dark Sector Mass Relations from RG Focusing}
\author{John Kearney and Aaron Pierce}
\vspace{0.2cm}
\affiliation{Michigan Center for Theoretical Physics (MCTP) \\
Department of Physics, University of Michigan, Ann Arbor, MI
48109}

\date{\today}

\begin{abstract}
Dark sector mass relations, such as those which permit near-threshold or near-resonance annihilation in the early universe, could arise due to IR-attractive ratios in renormalization group equations.
Achieving a particular ratio requires specific dark matter gauge charges or interactions, leading to predictions about the dark matter properties.  Furthermore, additional states with masses comparable to the dark matter mass may be necessary, potentially giving rise to novel phenomenology. 
We explore this idea in the context of dark matter charged under a new gauged $U(1)_X$ that kinetically mixes with the Standard Model hypercharge.
\end{abstract}

\maketitle

\section{Introduction}

In many models of electroweak-scale dark matter (DM), achieving the correct thermal relic density while avoiding direct and indirect detection, collider and precision constraints requires mass relations between particles in the dark sector. 
As first explored in \cite{Griest:1990kh}, the dark matter may be close in mass to another state, permitting coannihilation with or phase-space suppressed annihilation to the other state.
Alternatively, the dark matter mass may be approximately half that of a resonance.
Such relations can enhance the dark matter annihilation rate, allowing the correct relic density to be achieved with smaller couplings, and hence without large detection or production cross sections (see \cite{Cohen:2011ec,Profumo:2013hqa} for recent discussions).

But why should such mass relations exist?
Moreover, as masses and couplings vary with energy scale, one can ask why dark sector masses happen to exhibit the required relations at the appropriate scale (i.e. around the dark matter mass).
In this note, we explore the idea that dark sector mass relations arise from infrared (IR)-attractive ratios.  Though GUT scale parameters may be \emph{a priori} unrelated, renormalization group (RG) running focuses the parameters to particular ratios at the electroweak scale.
The mass relations thus emerge dynamically due to the interactions and quantum numbers of the dark sector particles.

\begin{figure}
\centering
\includegraphics[width=\linewidth]{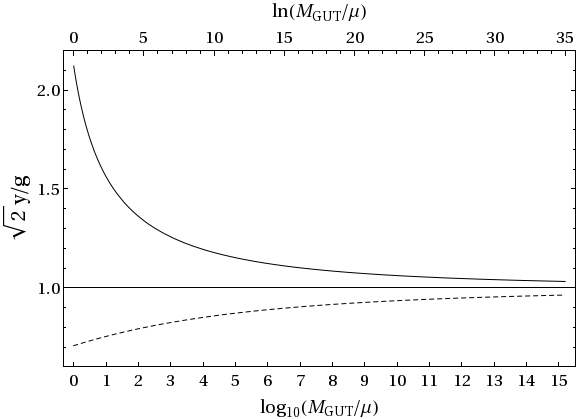}
\caption{\label{fig:intro_simplifiedfocusing} Evolution of the ratio $\sqrt{2} y/g$ as a function of scale $\mu$ in the simplified example of RG focusing based on \eqs{eq:simplegbeta}{eq:simpleybeta} with $c = 5, b = 1$ and $k = \frac{3}{2}$.
We fix $g_{GUT} = 2$ and take  $y_{GUT} = 3$ (solid) or $y_{GUT} = 1$ (dashed).}
\end{figure}

For instance, consider a fermion (which we imagine to be the DM) and a vector boson that both acquire mass via coupling to a scalar field that attains a vacuum expectation value (vev).
If $y$ represents the relevant Yukawa coupling, $g$ the gauge coupling and $V$ the vev, then the fermion and vector boson masses go as $m_f \propto y V$ and $m_V \propto g V$, such that the mass ratio $m_f/m_V \propto y/g$ is entirely determined by the ratio of the couplings.  At one-loop order, the RG equations for the couplings are of the form
\begin{eqnarray}
\label{eq:simplegbeta} (4 \pi)^2 \frac{d g}{d t} & = & b g^3, \\
\label{eq:simpleybeta} (4 \pi)^2 \frac{d y}{d t} & = & y (c y^2 - k g^2),
\end{eqnarray}
where $t \equiv \ln \mu$ is the logarithm of the renormalization scale $\mu$.  This system of equations exhibits an IR-attractive ratio, which can be found by solving
\begin{equation}
\label{eq:generalfixedratio}
\frac{d}{dt} \ln \left(\frac{y}{g}\right) = 0 \quad \Rightarrow \quad \left(\frac{y}{g}\right)_{IR} = \pm\sqrt{\frac{k + b}{c}}.
\end{equation}
Certain choices of quantum numbers and couplings (i.e. of $b,c$ and $k$) will lead to mass relations such as $m_f \approx m_V$ or $m_f \approx \frac{m_V}{2}$.
A toy example of the focusing of $\sqrt{2} y/g$ to the fixed ratio (of $1$) is shown in Fig.~\ref{fig:intro_simplifiedfocusing} for $c = 5, b = 1$ and $k = \frac{3}{2}$.  Clearly, a particular coupling (and hence mass) ratio can be achieved at the weak scale without significant numerical coincidence at the GUT scale.

This idea shares some intellectual ancestry with earlier attempts to predict masses and mass relations for the top quark and Higgs boson using IR fixed points in the Standard Model (SM) \cite{Pendleton:1980as,Hill:1980sq,Wetterich:1987az,Schrempp:1992zi}.
Other recent attempts to understand dark sector masses using RG properties include \cite{Hertzberg:2012zc,Hambye:2013dgv,Bai:2013xga}.

In the next section, we will explore RG focusing in the context of models in which the dark matter is charged under a new $U(1)_X$ gauge group that kinetically mixes with the hypercharge $U(1)_Y$ of the SM.  We will demonstrate how particular mass relations can be achieved and will discuss the phenomenological implications.  Then, we will discuss possible extensions and alternative applications of this idea and conclude.

\section{Kinetic Mixing Examples}
\label{sec:models}

A simple model of dark matter involves a fermion $\Psi$ charged under a new $U(1)_X$ gauge group,
\begin{equation}
\mathcal{L} \supset i \overline{\Psi} \gamma^\mu (\partial_\mu + i g_X (q_L P_L - q_R P_R) X_\mu) \Psi,
\end{equation}
where $X$ is the $U(1)_X$ gauge boson and $q_{L, R}$ are the $U(1)_X$ charges of the left- and right-handed components of $\Psi$.
The $X$ boson mixes with the Standard Model hypercharge boson $Y$ via kinetic mixing \cite{Holdom:1985ag,Baumgart:2009tn},
\begin{equation}
\mathcal{L} \supset - \frac{\sin \epsilon}{2} F_X^{\mu \nu} F_{Y \mu \nu}.
\end{equation}
We assume that $X$ acquires mass due to the vev of a scalar field $\Phi$ (with charge normalized to $-1$),
\begin{equation}
\mathcal{L} \supset \abs{D_\mu \Phi}^2 = \abs{(\partial_\mu - i g_X X_\mu) \Phi}^2,
\end{equation}
such that for $\vev{\Phi} = \frac{V}{\sqrt{2}}$, $m_X = g_X V$.  Diagonalizing the kinetic and mass terms gives rise to three mass eigenstates $(A, Z, Z^\prime)$, where $A$ is the SM photon and $(Z,  Z^\prime)$ are admixtures of the SM $Z$-boson and $X$.  This mixing allows the correct dark matter thermal relic density $\Omega h^2 = 0.1199 \pm 0.0027$ \cite{Ade:2013zuv} to be achieved, as $\Psi$ will annihilate to SM states via the $Z$ and $Z^\prime$ bosons.
Throughout this paper, we assume that the Higgs boson associated with the $U(1)_X$ breaking, $\varphi$, does not significantly impact the phenomenology.

\begin{figure}
\centering
\includegraphics[width=0.8\linewidth]{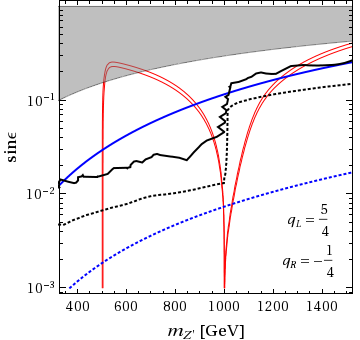}

\vspace{0.3cm}

\includegraphics[width=0.8\linewidth]{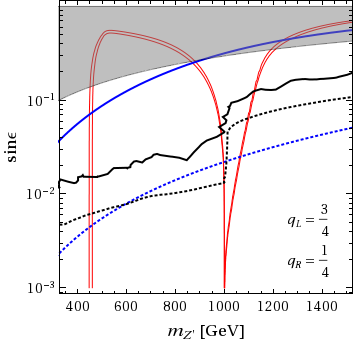}

\vspace{0.3cm}

\includegraphics[width=0.8\linewidth]{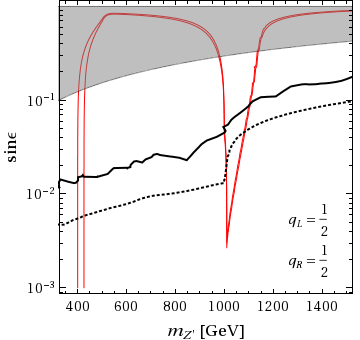}

\caption{\label{fig:sineps_mzp_constraints}
Regions in the $(m_{Z^\prime}, \sin \epsilon)$ plane yielding the correct relic density (red), taken to be the $5 \sigma$ range from PLANCK \cite{Ade:2013zuv}, fixing $m_{DM} = 500 \text{ GeV}$ and $g_X = 1$.
Also shown are constraints from the LHC (black, solid) \cite{CMS-PAS-EXO-12-061}, electroweak precision tests (gray shaded) \cite{Kumar:2006gm,Hook:2010tw} and XENON100 (blue, solid) \cite{Aprile:2012nq}.
In addition, we include projections for XENON1T (blue, dotted) \cite{Aprile:2009yh} and the 14 TeV LHC with $\mathcal{L} = 300 \text{ fb}^{-1}$ (black, dotted).
The three plots correspond to $q_L = \frac{5}{4}, q_R = -\frac{1}{4}$ (top), $q_L = \frac{3}{4}, q_R = \frac{1}{4}$ (middle) and $q_L = q_R = \frac{1}{2}$ (bottom); for $q_L = q_R = \frac{1}{2}$ the purely axial DM couplings yield velocity-suppressed SI scattering, so no XENON limits appear.}
\end{figure}

This type of model provides a particularly nice framework for studying RG focusing.
First, the relative simplicity permits the construction of straightforward yet instructive examples. 
Second, both theoretical and experimental considerations tend to require small $\sin \epsilon$, which makes it difficult to achieve the correct relic density without invoking particular mass relations \cite{Mambrini:2011dw}.
On the theoretical side, the value of $\sin \epsilon$ generated by loops of heavy particles charged under both $U(1)_X$ and $U(1)_Y$ is expected to be $\sin \epsilon \lsim 0.1$ \cite{Dienes:1996zr,Baumgart:2009tn}.
On the experimental side, LHC searches for resonances decaying to lepton pairs \cite{CMS-PAS-EXO-12-061} and electroweak precision measurements \cite{Kumar:2006gm,Hook:2010tw} place limits on $\sin \epsilon$ for a wide range of $m_{Z^\prime}$.  Moreover, if the dark matter exhibits vectorial couplings to $X$, direct detection constraints on spin-independent (SI) scattering with nucleons from XENON100 \cite{Aprile:2012nq} can be significant.\footnote{For weak-scale thermal dark matter, bounds from indirect detection experiments are not currently constraining \cite{Mambrini:2011dw}.  For lighter DM ($m_{DM} \lsim 10 \text{ GeV}$), limits from BaBar \cite{Aubert:2009af} can also be relevant \cite{Hook:2010tw}.}
The relevant experimental bounds, which tightly constrain $\sin \epsilon$, are shown in Fig.~\ref{fig:sineps_mzp_constraints}.\footnote{Relic densities and SI scattering cross sections are computed in \texttt{micrOMEGAs3.1} \cite{Belanger:2013oya} using expressions from \cite{Chun:2010ve}.  Approximate projections for the 14 TeV LHC with $\mathcal{L} = 300 \text{ fb}^{-1}$ are derived based on hadronic structure functions \cite{Carena:2004xs,Accomando:2010fz} calculated using \texttt{CalcHEP 3.4} \cite{Belyaev:2012qa}, dilepton invariant mass resolution estimates from \cite{Feldman:2006wb,Beringer:1900zz}, and variation in background between $\sqrt{s} = 8 \text{ TeV}$ and $14 \text{ TeV}$ estimated using \texttt{PYTHIA 8.1} \cite{Sjostrand:2007gs}.}

Consequently, for approximately weak-scale DM, achieving the correct thermal relic density with sufficiently small values of $\sin \epsilon$ requires either
\begin{enumerate}
\item $m_{DM} \approx m_{Z^\prime}$, such that the efficient annihilation process $\Psi \overline{\Psi} \rightarrow Z^\prime Z^\prime$ (which for $m_{DM} > m_{Z^\prime}$ would yield a very small relic density even if $\sin \epsilon \approx 0$) can occur, but Boltzmann and phase-space suppression prevent over-annihilation, or
\item $m_{\text{DM}} \approx \frac{1}{2} m_{Z^\prime}$, in which case annihilation $\Psi \overline{\Psi} \rightarrow Z^\prime \rightarrow \text{SM} \; \overline{\text{SM}}$ is enhanced in the early universe due to a small $s$-channel propagator, permitting smaller values of $\sin \epsilon$.
\end{enumerate}
The necessity of these mass relations makes kinetic mixing models with weak-scale DM prime candidates for benefitting from RG focusing.
We now present two models with basic structure as outlined in the introduction, one of which exhibits $m_f \approx m_{Z^\prime}$ and one of which exhibits $m_f \approx \frac{1}{2} m_{Z^\prime}$.

\subsection{(1) $m_{DM} \approx m_{Z^\prime}$}

Consider $\chi_\pm$, $\eta_\pm$ to be left-handed Weyl fermions with $U(1)_X$ charges $\pm q$ and $\pm (1-q)$ respectively.  We introduce Yukawa couplings of the form
\begin{equation}
\label{eq:model1}
\mathcal{L} \supset - y_+ \Phi \chi_+ \eta_+ - y_- \Phi^\ast \chi_- \eta_- + \text{h.c.}
\end{equation}
As the fermions come in pairs with opposite charges, this model is anomaly free.
We assume separate $\mathbb{Z}_2$ symmetries, which ensure that the new fermions are stable (and hence DM candidates) and also forbid vector-like masses of the form $\chi_+ \chi_-$.  
After spontaneous symmetry breaking of the $U(1)_X$, the $\chi_\pm$ and $\eta_\pm$ are married to yield two Dirac fermions with masses $m_\pm = \frac{y_\pm V}{\sqrt{2}}$.  The ratio of $m_\pm$ to $m_X$ is given by
\begin{equation}
\label{eq:m1massratio}
\frac{m_\pm}{m_X} = \frac{y_\pm}{\sqrt{2} g_X}.
\end{equation}
The one-loop beta functions for the couplings are
\begin{eqnarray}
\label{eq:m1yukawabeta}
(4 \pi)^2 \frac{d y_\pm}{d t} & = & y_\pm \left(2 y_\pm^2 + y_\mp^2 - 3 (q^2 + (1-q)^2) g_X^2\right), \\
\label{eq:m1gaugebeta}
(4 \pi)^2 \frac{d g_X}{d t} & = & b_X g_X^3,
\end{eqnarray}
where $b_X = \frac{4}{3} (q^2 + (1-q)^2) + \frac{1}{3}$.  This system of equations exhibits IR-attractive fixed ratios
\begin{equation}
\label{eq:m1couplingratio}
\left.\frac{y_+}{y_-}\right|_0 = 1, \quad \left.\frac{y_\pm}{g_X}\right|_0 = 
\frac{1}{3} \sqrt{13 (q^2 + (1-q)^2) + 1}.
\end{equation}
The subscript ``$0$'' denotes that these ratios are RG invariant -- in other words, for couplings fixed to these ratios, the ratios will be preserved by RG running.

We now imagine that the couplings take some generic values at the unification scale $M_{GUT}$.  Then, as the couplings are run to the dark matter scale (taken to be on the order of $m_Z$), they evolve such that they are attracted towards these ratios.  By \eq{eq:m1massratio}, this leads to particular relations between the fermion masses and the $Z^\prime$ mass -- different choices of $q$ will yield different mass ratios.

By examining \eqs{eq:m1massratio}{eq:m1couplingratio}, we see that we can approximately achieve the desired mass relation if $q = \frac{5}{4}$, for which
\begin{equation}
\left.\frac{m_\pm}{m_X}\right|_0 = \left.\frac{y_\pm}{\sqrt{2} g_X}\right|_0 \approx 1.1.
\end{equation}
Provided that the couplings converge to this ratio sufficiently quickly, the dark matter will have mass $m_\chi \gsim m_{Z^\prime}$ and so will undergo phase-space suppressed annihilation to $Z^\prime Z^\prime$ in the early universe, conceivably yielding the correct relic density even for very small values of $\sin \epsilon$.
Both of the new fermions are stable, so they will each constitute a component of the dark matter -- however, the heavier state will annihilate more efficiently and so the lighter state will comprise the majority of the dark matter.

How quickly do the couplings converge to the fixed ratio?  Consider the variable $\delta_\pm$, defined by
\begin{equation}
\frac{y_\pm}{g_X} = \left(\frac{y_\pm}{g_X}\right)_0 (1 + \delta_\pm),
\end{equation}
which measures the deviation of the coupling ratio from the fixed ratio.  From \eqs{eq:m1yukawabeta}{eq:m1gaugebeta}, we can derive a differential equation for $\delta_\pm$, assuming $\delta_+ = \delta_-$ for simplicity\footnote{As this is a point of enhanced symmetry, $y_+ = y_-$ could perhaps be enforced as a GUT scale relation.},
\begin{equation}
\label{eq:m1delta}
\frac{d \delta_\pm}{dt} = \frac{3 g_X^2}{(4 \pi)^2} \left(\frac{y_\pm}{g_X}\right)^2_0 \delta_\pm (\delta_\pm + 1) (\delta_\pm + 2).
\end{equation}
This demonstrates that, for $\delta_\pm > 0$ or $-1 < \delta_\pm < 0$, $\delta_\pm \rightarrow 0$ as $t \rightarrow -\infty$; the fixed ratio is IR attractive.  The other fixed points of the equation are $\delta_\pm = -1$, corresponding to turning off the Yukawas (and indicating no Yukawas are generated by RG running), and $\delta_\pm = -2$, which is analogous to the fixed point at $\delta_\pm = 0$ up to re-phasing of the fermion fields.

\begin{figure*}
\centering
\subfigure{\includegraphics[width=0.46\linewidth]{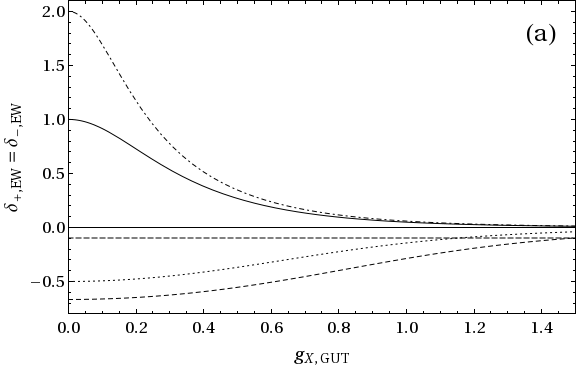}} \hspace{0.05\linewidth}
\subfigure{\includegraphics[width=0.46\linewidth]{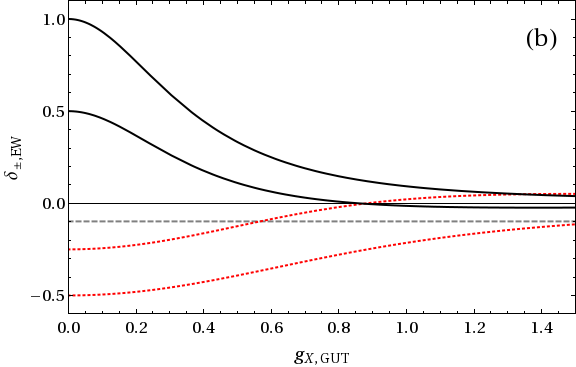}}
\caption{\label{fig:threshfoceff} 
$\delta_\pm$ at the electroweak scale after 33 $e$-folds of RG evolution as a function of $g_{X, GUT}$ assuming (a) $\delta_+ = \delta_-$ and (b) $\delta_+ \neq \delta_-$.
Black lines in (a) represent $\delta_{\pm, GUT} = 2$ (dot-dashed), $1$ (solid), $-1/2$ (dotted) and $-2/3$ (dashed).
In (b), we take $(\delta_+, \delta_-)_{GUT} = (1/2,1)$ (black) and $(\delta_+, \delta_-)_{GUT} = (-1/4,-1/2)$ (red, dotted).  In both plots, the gray dashed line represents the value of $\delta_{\pm, EW}$ for which $m_\pm = m_X$.}
\end{figure*}

The values of $\delta_{+, EW} = \delta_{-, EW}$ at the electroweak scale after $\sim 33$ $e$-folds of running 
(corresponding to running from $\mu = M_{GUT}$ to $\mu \sim \mathcal{O}(m_Z)$\footnote{As the gauge couplings do not unify in this minimal model, it is not obvious what value one should take for $M_{GUT}$.  
Potential candidates range from the scale at which $g_1$ and $g_2$ unify all the way up to the Planck scale, and depend on the UV completion.  
We remain agnostic, and simply take $\ln(M_{GUT}/m_{DM}) \approx 33$ as a representative value.}) 
are shown in Fig.~\ref{fig:threshfoceff}(a) as a function of $g_{X, GUT}$ for a variety of GUT-scale deviations $\delta_{+, GUT} = \delta_{-, GUT}$.  It is clear that, for reasonable values of $g_{X, GUT} \approx \mathcal{O}(1)$, the couplings come very close to the fixed ratio even if there is significant misalignment at the GUT scale, demonstrating the efficacy of the focusing.  Thus, this mechanism is capable of generating dark sector mass relations without substantial coincidence of parameters.  As expected from \eq{eq:m1delta}, the couplings approach the fixed ratio faster for $\delta_\pm > 0$ than for $-1 < \delta_\pm < 0$.

\begin{figure}
\centering
\includegraphics[width=\linewidth]{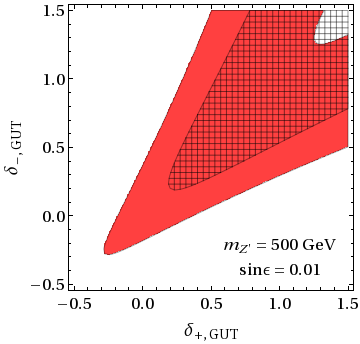}
\caption{\label{fig:correctrelic_d1d2} Values of $\delta_{\pm, GUT}$ yielding the correct relic density for $g_{X, GUT} = 1.2$ (hatched) or 1.4 (red).  We fix $m_{Z^\prime} = 500 \text{ GeV}$ and $\sin \epsilon = 0.01$.  For $\sin \epsilon \lsim 0.015$ (chosen to satisfy the LHC limit shown in the top panel of Fig.~\ref{fig:sineps_mzp_constraints} -- here $q_{L,\pm} = \pm q = \pm \frac{5}{4}$ and $q_{R,\pm} = \pm(1-q) = \mp\frac{1}{4}$), the precise value of $\sin \epsilon$ does not affect the cosmology provided that it is large enough that the $Z^\prime$ decays prior to BBN.}
\end{figure}

It is also interesting to consider what happens if the Yukawa couplings are not aligned at the GUT scale ($\delta_{+, GUT} \neq \delta_{-, GUT}$).  The results are shown in Fig.~\ref{fig:threshfoceff}(b).  Although the Yukawas do not end up equal, they are driven to similar values near the IR-attractive ratio.  This gives rise to the situation described above wherein the dark matter is multi-component, but dominated by the (slightly) lighter component.  In Fig.~\ref{fig:correctrelic_d1d2}, we show the regions in the $(\delta_+, \delta_-)_{GUT}$ plane for which the correct relic density is achieved for two different choices of $g_{X, GUT}$.
As a result of the RG focusing, a significant region of the GUT scale parameter space yields the correct relic density.

Our analysis has thus far considered the RG evolution of the couplings only at one-loop.  Given the large GUT scale values for the couplings, a reasonable concern is whether our conclusions are greatly affected by higher-order terms.  
For instance, in Fig.~\ref{fig:threshfoceff}(a), $y_{\pm, GUT} = 7.0$ for $\delta_{\pm, GUT} = 2.0$ and $g_{X, GUT} = 1.5$, so in this region the plot should be taken as indicative of the power of one-loop focusing as opposed to an exact result.
Performing a full analysis of higher-loop effects is more complicated, in part because the (so far unspecified) scalar quartic coupling enters at the two-loop level.  However, we have confirmed that higher-loop corrections of the size expected from \cite{Machacek:1983fi,Luo:2002ti} do not significantly alter our results or the rate of convergence to the fixed ratio.  This is partly because the couplings become smaller in the IR, such that the perturbative expansion is under control in the region where the couplings are approaching the fixed ratio.  As a result, the one-loop terms dominate.

\begin{figure}
\centering
\includegraphics[width=\linewidth]{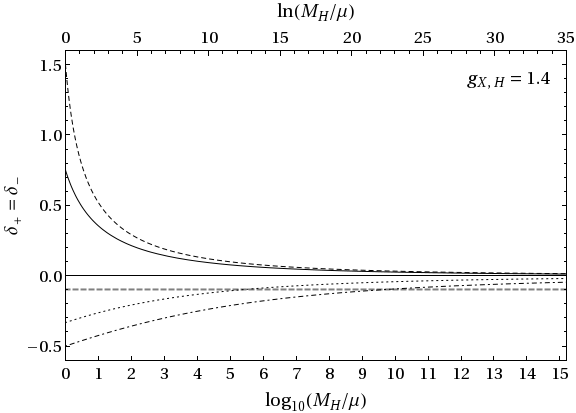}
\caption{\label{fig:focus_func_efolds} $\delta_+ = \delta_-$ as a function of $\log(M_H/\mu)$ for $g_{X, H} = 1.4$ (the gauge coupling at $\mu = M_H$).  The gray dashed line represents the value of $\delta_{\pm, EW}$ for which $m_\pm = m_X$.}
\end{figure}

Finally, it is interesting to explore how efficient the focusing would be over fewer $e$-folds.  For instance, one could imagine a scenario in which the RG equations attain the correct form to yield the desired IR-attractive ratio after crossing some heavy mass threshold $M_H < M_{GUT}$.  In this case, we can take $(\delta_\pm, g_X)_H$ to be boundary conditions at the threshold scale $\mu = M_{H}$.   In Fig.~\ref{fig:focus_func_efolds}, we show how $\delta_+ = \delta_-$ evolves as a function of $\ln(M_{H}/\mu)$, fixing $g_{X, H} = 1.4$.  It is evident that $\delta_+ = \delta_-$ approaches zero quite rapidly, particularly for $\delta_{\pm, H} > 0$.  Even for $\delta_{+, H} = \delta_{-, H} = 1.5$, $\delta_+ = \delta_- \lsim 0.05$ by $\log_{10}(M_{H}/\mu) = 8$.  Thus, this mechanism could be used to generate mass relations in models with mass thresholds as low as $M_{H} \approx 10^{10} \text{ GeV}$.
In the context of kinetic mixing models, this mass threshold could perhaps correspond to the mass of heavy states charged under $U(1)_X$ and $U(1)_Y$ responsible for generating $\sin \epsilon$.\footnote{The rapidity of the focusing also implies that such a model could give rise to dark sector mass relations at a significantly higher scale than the weak scale -- of course, such a scenario is phenomenologically dismal.}

The main phenomenological implication of this model is that the dark matter $U(1)_X$ charges must be $q_L \approx \frac{5}{4}$, $q_R \approx - \frac{1}{4}$ in order to achieve the desired mass ratio.
For these charge assignments, the dark matter exhibits a significant vectorial coupling to the $Z$ and $Z^\prime$ gauge bosons, giving rise to appreciable SI scattering cross sections and enabling direct  detection experiments to probe smaller values of $\sin \epsilon$.
Depending on the value of $m_\pm$, the strongest constraints in the near threshold region come from either LHC dilepton resonance searches or XENON100 and require $\sin \epsilon \lsim (1-2) \times 10^{-2}$ (see the top panel of Fig.~\ref{fig:sineps_mzp_constraints}).  The DM could likely be observed by a one-ton Xenon experiment for $\sin \epsilon \gsim 10^{-3}$, and for $\sin \epsilon \gsim 5 \times 10^{-3}$ the concurrent observation of the DM and a $Z^\prime$ with $m_{Z^\prime} \approx m_{DM}$ may be possible.

\subsection{(2) $m_{DM} \approx \frac{1}{2} m_{Z^\prime}$}
  
As shown in Fig.~\ref{fig:sineps_mzp_constraints}, the other region of interest exhibiting the correct relic density and small $\sin \epsilon$ has $m_{DM}/m_{Z^\prime} \approx \frac{1}{2}$, such that annihilation in the early universe is approximately on resonance.  Thus, one might also wish to explain this mass ratio via a similar mechanism.
However, for the model above,
\begin{equation}
\left.\frac{2 m_\pm}{m_X}\right|_0 = \left.\frac{\sqrt{2} y_\pm}{g_X}\right|_0 \geq 1.3
\end{equation}
with the minimum occurring for $q = \frac{1}{2}$.  Thus, we are limited in how close we can get to $m_{DM}/m_{Z^\prime} \approx \frac{1}{2}$, at least in this simple model.
Gauge couplings drive $y_\pm$ up towards the IR, whereas Yukawas drive $y_\pm$ down, so to achieve $y_\pm$ sufficiently small with respect to $g_X$ requires the introduction of additional Yukawa couplings.
Said another way, in terms of \eq{eq:generalfixedratio}, to get closer to resonance requires additional Yukawa contributions that increase $c$ without a correspondingly large increase in $b$ ($k$ is fixed by the DM charges).

This can be accomplished by introducing new fermions with Yukawa couplings to $\Phi$.  The additional fermions will contribute to the scalar wave function renormalization, increasing $c$.\footnote{A similar alternative, which we do not elaborate on here, would be to introduce new ``inert'' scalars coupling to $\chi_\pm, \eta_\pm$, which would increase $c$ by contributing to the fermion wave function renormalization.}
Moreover, if these states have larger $U(1)_X$ charges than the dark matter or are charged under additional gauge groups, their Yukawa couplings will tend to larger values than the DM Yukawa couplings, further enhancing $c$ and making it easier to achieve the ratio $m_{DM}/m_{Z^\prime} \approx \frac{1}{2}$.
However, the introduction of additional couplings can somewhat reduce the efficacy of the focusing relative to the $m_{DM} \approx m_{Z^\prime}$ case above.

Perhaps the simplest way to introduce new states is to augment \eq{eq:model1} to respect an $SU(N_F)^2$ symmetry.  For $N_F = 4$ and $q = \frac{1}{2}$, $(2m_\pm/m_X)_0 = (\sqrt{2} y_\pm/g_X)_0 \approx 1.0$.  However, as there are more DM components, the dark matter must annihilate more efficiently to achieve the correct relic density.
This requires either a larger value of $\sin \epsilon$ (in tension with the constraints mentioned above) or that $2 m_\pm$ is particularly close to $m_{Z^\prime}$, which would imply a significant numerical coincidence in GUT scale parameters even with RG focusing (largely neutralizing the benefits of the focusing).

Consequently, we instead introduce new fermions $X_\pm, N_\pm$ (in addition to $\chi_\pm, \eta_\pm$) that couple to $\Phi$, but decay such that they do not contribute to the dark matter relic density.  $X_\pm$ and $N_\pm$ have $U(1)_X$ and $U(1)_Y$ charges $\pm Q_X, \pm (1-Q_X)$ and $\pm Q_Y, \mp Q_Y$ respectively.  We add to \eq{eq:model1} the Yukawa couplings
\begin{equation}
\mathcal{L} \supset - Y_+ \Phi X_+ N_+ - Y_- \Phi^\ast X_- N_- + \text{h.c.}
\end{equation}
When $\Phi$ takes on its vev, $X_\pm$ and $N_\pm$ marry to form two Dirac fermions with $M_\pm = \frac{Y_\pm V}{\sqrt{2}}$.  As the $X_\pm, N_\pm$ states decay, in principle we do not need to relate their masses to that of the $U(1)_X$ gauge boson as for $\chi_\pm, \eta_\pm$.
However, since the interactions of the new states are vital for producing the desired IR-attractive ratio, we want these states to contribute to the RG evolution all the way to the dark scale.  
In light of this, it is logical that these states acquire all of their mass from $U(1)_X$ breaking such that $M_\pm \sim m_\pm$ -- for this reason, we assume additional $\mathbb{Z}_2$ symmetries forbidding vector-like mass terms.  This also leads to a particular prediction of these models, namely the existence of additional dark sector states 
with masses comparable to the dark matter mass.

The choices of $Q_X$ and $Q_Y$ determine how $X_\pm, N_\pm$ can decay -- one choice that readily permits decay is $Q_X = q$ and $Q_Y = 1$.  We introduce a new scalar $\tilde{e}$ with $SU(2)_L \times U(1)_Y$ quantum numbers $(\mathbf{1},-1)$ and interactions of the form
\begin{equation}
- \Delta \mathcal{L} = \kappa_+ \tilde{e} X_+ \chi_- + \kappa_- \tilde{e} N_- \eta_+ + \kappa \tilde{e}^\dagger \ell \ell + \text{h.c.},
\end{equation}
permitting decays such as $X^- \rightarrow \chi \ell^- \overline{\nu}_\ell$ (assuming $m_{\tilde{e}} > M_\pm > m_\pm$ -- superscripts denote $U(1)_{\text{EM}}$ charges).\footnote{Another choice permitting decay is $Q_X = 1$, $Q_Y = 0$.  The $N_\pm$ would be gauge singlets, and could decay via the higher-dimension operator $\mathcal{L} = \frac{1}{\Lambda} N_\pm u^c d^c d^c$.  The attractive ratio in this model is $(2m_\pm/m_X)_0 \approx 1.0$ for $q = \frac{4}{5}$.}
Note that the $U(1)_Y$ interactions will tend to drive $Y_\pm > y_\pm \Rightarrow M_\pm > m_\pm$.  $\kappa_\pm$ and $\kappa$ are taken to be sufficiently small that they have a negligible effect on the dark sector RG evolution, but sufficiently large that the $X_\pm, N_\pm$ states decay prior to DM freeze-out to avoid repopulating the dark matter.  For approximately TeV scale particles, fast enough decay occurs if $\kappa_\pm \approx \kappa \gsim 10^{-4}$ such that both of these conditions can indeed be satisfied.

\begin{figure}
\centering
\includegraphics[width=\linewidth]{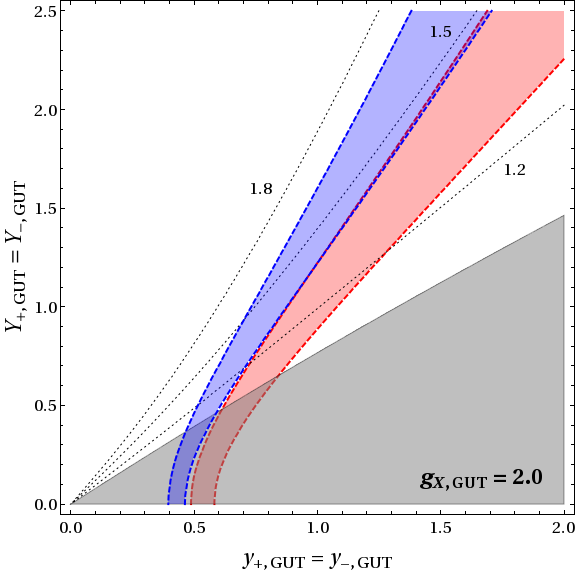}
\caption{\label{fig:resfoc_yY_g2} Regions in the $(y_\pm, Y_\pm)_{GUT}$ plane for which $2m_\pm/m_X \in [0.95,1.05]$ (left- and right-boundaries, respectively) for $q = \frac{3}{4}$ (blue) and $q = \frac{1}{2}$ (red), fixing $g_{X, GUT} = 2$.  The dotted contours give the value of $M_\pm/m_\pm$ at the weak scale for $q = \frac{1}{2}$, with the shaded gray region forbidden as $M_\pm < m_\pm$ -- contours for $q = \frac{3}{4}$ are not shown but are largely similar.}
\end{figure}

In this model, the ratios that the couplings approach in the IR are somewhat more complicated due to the effect of the hypercharge on the RG equations.
Symmetry between $+$ and $-$ states implies
\begin{equation}
\left.\frac{y_+}{y_-}\right|_0 = \left.\frac{Y_+}{Y_-}\right|_0 = 1.
\end{equation}
However, solving the equations
\begin{equation}
\frac{d}{dt} \ln\left(\frac{y_\pm}{g_X}\right) = 0, \quad \frac{d}{dt} \ln\left(\frac{Y_\pm}{g_X}\right) = 0
\end{equation}
yields
\begin{eqnarray}
\left.\frac{y_\pm}{g_X}\right|_0 & = & \sqrt{\frac{17 (q^2 + (1-q)^2) + 1 - 36 (g_Y/g_X)^2}{15}}, \\
\left.\frac{Y_\pm}{g_X}\right|_0 & = & \sqrt{\frac{17 (q^2 + (1-q)^2) + 1 + 54 (g_Y/g_X)^2}{15}}.
\end{eqnarray}
The attractive ratios evolve as a function of scale (or as a function of the values of $g_{X, Y}$).
Fig.~\ref{fig:resfoc_yY_g2} shows regions of GUT parameter space for which $\frac{2 m_\pm}{m_X} \in [0.95,1.05]$ at the weak scale for two charge assignments $q = \frac{3}{4}$ and $q = \frac{1}{2}$, taking $g_{X, GUT} = 2$.  For simplicity, we set $y_+ = y_-$ and $Y_+ = Y_-$.
The chosen range for $2m_\pm/m_X$ provides a rough guideline as to where the correct thermal relic density is achieved, consistent with experimental constraints, for dark matter masses $\mathcal{O}(100 \text{ GeV} - 1 \text{ TeV})$ and $\sin \epsilon \lsim \mathcal{O}(0.1)$.
However, valid regions of parameter space do exist for smaller or larger values of $2m_\pm/m_X$.

\begin{figure}
\centering
\includegraphics[width=\linewidth]{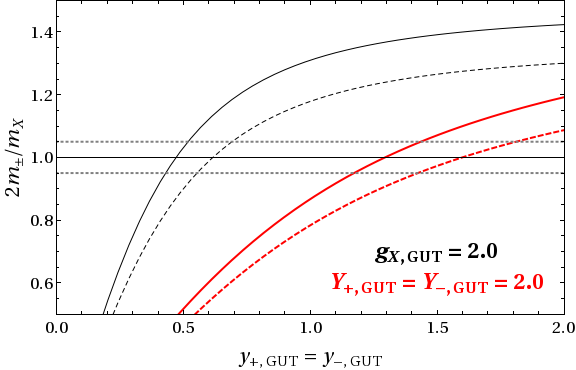}
\caption{\label{fig:resfoc_y_g2_Y2} The distance from resonance at the weak scale (parameterized by $2 m_\pm/m_X$), fixing $g_{X, GUT} = 2$, as a function of $y_{+, GUT} = y_{-, GUT}$ for $q = \frac{3}{4}$ (solid) and $q = \frac{1}{2}$ (dashed) in the model without (black) and with (red) the $X_\pm, N_\pm$ states and $Y_{+, GUT} = Y_{-, GUT} = 2$.  The presence of the extra states with reasonable GUT-scale Yukawas reduces the numerical coincidence required to achieve $m_{\pm} \approx \frac{1}{2} m_X$.  Gray dotted lines demarcate the region $2 m_\pm/m_X \in [0.95,1.05]$.}
\end{figure}

In Fig.~\ref{fig:resfoc_y_g2_Y2}, we show the value of $2 m_{\pm}/m_X$ at the weak scale as a function of $y_{+, GUT} = y_{-, GUT}$ both with and without the $X_\pm, N_\pm$ states.
If these states are present, the slope of the lines is much shallower in the region of $2m_\pm/m_X = 1$, such that a wider range of $y_{+, GUT} = y_{-, GUT}$ will give rise to $\frac{2 m_\pm}{m_X} \in [0.95,1.05]$.
Without the additional states, a more significant conspiracy of GUT scale parameters is needed to achieve $m_\pm \approx \frac{1}{2} m_X$.

Again, these results are based on one-loop beta functions only, neglecting the small kinetic mixing, but we have checked that approximate corrections due to two-loop effects and kinetic mixing \cite{delAguila:1988jz,Luo:2002iq} do not significantly alter our results.  However, because the spectrum contains states charged under both $U(1)_X$ and $U(1)_Y$, a related consideration is how $\sin \epsilon$ evolves.  In particular, one might wonder what values of $(\sin \epsilon)_{GUT}$ yield the desired $(\sin \epsilon)_{EW} \sim \mathcal{O}(0.1)$.  Generally, depending on the precise choices of $q$ and $(\sin \epsilon)_{EW}$, either $(\sin \epsilon)_{GUT} \sim \mathcal{O}(0.01)$ or $(\sin \epsilon)_{GUT} \sim \mathcal{O}(0.5)$ for $g_{X, GUT} = 2$.  $(\sin \epsilon)_{GUT}$ is expected to be $\mathcal{O}(1)$ if the operator $F_X F_Y$ is permitted at the GUT scale or $\sim 0$ if it is forbidden (by, e.g., gauge invariance of the unification group).  Notably, the GUT boundary conditions required to give $(\sin \epsilon)_{EW} \sim \mathcal{O}(0.1)$ in this model are approximately consistent with one of these two scenarios.

If the dark matter relic density is set by near-resonant annihilation in the early universe, it may well imply the existence of additional states close in mass to the dark matter, resulting in novel phenomenology beyond the dark matter direct detection prospects.  For the model above, we predict new charged particles with masses $M_\pm \sim (1.2-1.7) m_\pm$.  As these particles decay prior to dark matter freeze-out, their lifetimes satisfy
\begin{equation}
\tau \lsim H^{-1}(T_{fo}) \; \Rightarrow \; \tau \lsim 10^{-9} \left(\frac{500 \text{ GeV}}{m_\pm}\right)^2 \text{ s}.
\end{equation}
For $\tau$ close to saturating this bound, the additional particles would be relatively long-lived and could produce disappearing tracks at the LHC.  Such signals have been searched for, and limits of $M_\pm \gsim \mathcal{O}(400-500 \text{ GeV})$ have been placed \cite{ATLAS-CONF-2013-069}.  For shorter lifetimes, the heavier states will decay to yield opposite-sign dilepton plus missing energy signatures, such that they could potentially be observed in SUSY chargino searches \cite{ATLAS-CONF-2013-049,CMS-PAS-SUS-13-006}.  However, the current reach of such searches is relatively limited (only requiring $M_\pm \gsim \mathcal{O}(100-200 \text{ GeV})$) due to the somewhat small $M_\pm-m_\pm$ splitting predicted.

Depending on the $U(1)_X$ charge of the dark matter, there can be interesting interplay between direct detection and LHC dilepton resonance searches.  For instance, if $q = \frac{3}{4}$, there are regions of parameter space exhibiting the correct relic density that are not yet excluded by current constraints, but which will be probed by both XENON1T and the LHC with $\sqrt{s} = 14 \text{ TeV}$ and $\mathcal{L} = 300 \text{ fb}^{-1}$.  These regions present the exciting possibility of the concurrent observation of the dark matter, a $Z^\prime$ boson with $m_{Z^\prime} \approx 2 m_{DM}$, and a long-lived charged particle with mass $m_{DM} < M_\pm < m_{Z^\prime}$.  In other regions of parameter space (or for $q = \frac{1}{2}$) the dark matter will evade direct detection, but this could be mitigated by the imminent observation of a $Z^\prime$ and perhaps also of a long-lived charged particle with mass $\frac{m_{Z^\prime}}{2} < M_\pm < m_{Z^\prime}$.  Of course, for $2 m_{DM}$ very close to $m_{Z^\prime}$, the small values of $\sin \epsilon$ that yield the correct relic density preclude both dark matter direct detection and $Z^\prime$ observation.  The charged states may still be observed, although this would of course depend on their masses and lifetimes.

\section{Discussion and Conclusions}
\label{sec:conc}

In this paper, we have proposed that dark sector mass relations may arise due to IR-attractive ratios in the dark sector RG equations.  We have discussed this in the context of two simple models consisting of new dark sector fermions charged under a gauged $U(1)_X$, which kinetically mixes with the SM $U(1)_Y$.  

We have focused on this class of model in part because it permits a straightforward introduction to this application of RG focusing, but a wide variety of alternative implementations can be imagined.  Throughout this paper, we have assumed that the Higgs boson $\varphi$ associated with the $U(1)_X$ breaking by the vev of $\Phi$ ($V$) does not impact the phenomenology.  However, the mass of $\varphi$ will also be related to $V$ by $m_\varphi \sim \sqrt{\lambda} V$, where $\lambda$ represents the $\Phi$ quartic coupling.  RG focusing could yield $m_{DM} \approx \frac{1}{2} m_\varphi$ -- in the presence of a mixed $\Phi$, SM Higgs quartic $\lambda_{\Phi H} \abs{\Phi}^2 \abs{H}^2$, this would lead to a realization of resonant Higgs portal dark matter \cite{LopezHonorez:2012kv}.
The fact that $b_X > 0$ for an Abelian gauge group meant that achieving $m_{DM} \approx \frac{1}{2} m_X$ required additional Yukawa couplings.  A non-Abelian theory could potentially allow this fixed ratio to be achieved more readily.  Of course, in this case, it is less trivial to communicate between the dark and SM sectors as gauge invariance now forbids kinetic mixing terms.
Alternatively, one could construct a coannihilation model in which $(M_\pm/m_\pm)_0 \gsim 1$, for instance if the heavier states had additional gauge interactions.  This is somewhat similar to the small mass splittings between charged and neutral states within multiplets that arise from electromagnetic interactions.
Aside from the new model-building possibilities, it may also be the case that pre-existing models of weak-scale dark matter exhibit RG focusing.

We have also made several simplifying assumptions regarding the structure of the theory.
For instance, we assumed no dynamics affect the RG equations between $\mu = M_{GUT}$ and $\mu \sim \mathcal{O}(m_Z)$.  As alluded to earlier, however, one could imagine more complicated scenarios with additional mass thresholds that alter the relative running of the couplings.
Furthermore, we considered only a single $U(1)_X$ Higgs field -- in models with multiple Higgs fields, the presence of additional ``$\tan \beta$'' parameters will affect the masses of the various particles.  While this provides more model-building freedom, it requires an explanation as to why ratios of vevs would take particular values as well.
In addition, we have remained agnostic as to why the dark scale and electroweak scale might be related (in other words, why $V \approx v_{EW}$).
This represents a second hierarchy problem, and could be addressed in a UV-complete model.  On a related note, one could attempt to realize a supersymmetric version of this mechanism.  In practice, the additional states present in a supersymmetric theory (which contribute to $b$, must acquire masses etc.) make such examples more complicated.

In models that generate mass relations via RG focusing, achieving specific attractive ratios requires certain charge assignments or the introduction of additional states and interactions.  While new states need not contribute to the dark matter relic density, they may still have properties (such as masses or charges) related to the dark matter properties.  Thus, although dark sector mass relations may make direct detection more difficult, they may also point to rich alternative phenomenology.  
Furthermore, RG focusing can make tightly-constrained models (such as the kinetic mixing models explored here) more palatable by relating masses without requiring serious numerical coincidences.
RG focusing in the dark sector offers a wide variety of possibilities and is worthy of future study.

\vspace{-0.2cm}
\subsection*{Acknowledgments}
\vspace{-0.3cm}
We would like to thank Kathryn Zurek, Michele Papucci and James Wells for useful conversations. AP
is supported by DoE grant DE-SC0007859 and CAREER grant NSF-PHY 0743315. JK is supported by CAREER grant NSF-PHY 0743315.

\bibliography{DarkSectorRG.bib}
\end{document}